\documentclass[aps,prd,showkeys,showpacs,amssymb,cite,
amsfonts,epsf,preprintnumbers,nofootinbib
]{revtex4}

\usepackage[dvips]{graphicx}
\usepackage{bm,latexsym,amsmath,amssymb,amsfonts}
\usepackage[usenames,dvipsnames]{color}
\usepackage[colorlinks=true,linkcolor=blue]{hyperref}
\usepackage{color}
\usepackage{soul}
\newcommand{\be}[1]{\begin{equation} \label{#1}}
\newcommand{\ee}{\end{equation}}
\newcommand{\bear}{\begin{eqnarray}}
\newcommand{\eear}{\end{eqnarray}}
\newcommand{\ba}{\begin{array}}
\newcommand{\ea}{\end{array}}

\newcommand{\G}{\mathcal{G}}

\begin{document}
\title{Equation of State in the Presence of Gravity}
\author{Hyeong-Chan Kim}
\email{hyeongchan@gmail.com}
\affiliation{Department of Physics, North Carolina State University, Raleigh, NC 27695-8202,}
\affiliation{School of Liberal Arts and Sciences, Korea National University of Transportation, Chungju 380-702, Korea}
\author{Gungwon Kang}
\email{gwkang@kisti.re.kr}
\affiliation{Korea Institute of Science
and Technology Information (KISTI), 245 Daehak-ro, Yuseong-gu,
Daejeon 34141, Korea }

\date[]{}

%
\begin{abstract}
We investigate how an equation of state for matter is affected when a gravity is present.
For this purpose, we consider a box of ideal gas in the presence of Newtonian gravity.
In addition to the ordinary thermodynamic quantities, a characteristic variable that represents a weight per unit area relative to the average pressure is required in order to describe a macroscopic state of the gas.
Although the density and the pressure are not uniform due to the presence of gravity, the ideal gas law itself is satisfied for the thermodynamic quantities when averaged over the system.
Assuming that the system follows an adiabatic process further, we obtain a {\it new} relation between the averaged pressure and density, which differs from the conventional equation of state for the ideal gas in the absence of gravity.
Applying our results to a small volume in a Newtonian star, however,
we find that the conventional one is reliable for most astrophysical situations when the characteristic scale is small.
On the other hand, gravity effects become significant near the surface of a Newtonian star.
\end{abstract}
\maketitle


\section{Introduction}

An equation of state (EOS) is a thermodynamic relation describing the state of matter under a given set of physical conditions.
Various EOSs are necessary to describe systems with strong gravity such as the sun,  white dwarf stars, neutron stars~\cite{Lattimer:2012nd} and the early universe~\cite{cosmology}.
EOSs used in these cases are usually obtained assuming the absence of gravity although gravity  is essential in all those systems.
In fact, the action of gravity modifies the distribution of particles.
Therefore, an examination of whether or not the EOS will be modified due to the presence of gravity is interesting.

As a related work, a modification of the relation between the pressure and the volume was studied by Bonnor~\cite{Bonnor} for a self-gravitating spherical mass of isothermal ideal gas. It, however, focused mainly on the issue of gravitational instability, which was recently re-examined by Lombardi and Bertin~\cite{Lombardi:2001ms}, without imposing any specific shape or symmetry to the self-gravitating object. Gravitational effects on the EOS for nuclear matter in neutron stars were also studied in the context of a solitonic Skyrme model~\cite{adam}.
Recently, a clue suggesting a possible modification of the local EOS due to the effect of gravity was given~\cite{Kim:2013nna} in the context of the Eddington-inspired Born-Infeld gravity~\cite{Banados:2010ix}.
Gravity theories, such as the Eddington-inspired Born-Infeld gravity and the Palatini $f(R)$ gravity~\cite{Barausse:2007ys,Sotiriou:2008rp,Olmo:2011uz,Pani:2012qd}, were known to have an inherent deficit, {\it e.g.}, the so-called surface singularity problem.
Namely, a curvature singularity tends to develop at the surface of polytropic stars.
In Ref.~\cite{Kim:2013nna}, the author showed that the problem could be resolved when the effect of strong gravity (or curvature) on EOS was taken into account.
One may ask whether similar effects of gravity on the EOS should be taken into account even in general relativity and Newtonian gravity.
In order to get some hints about this issue, we consider a box of ideal gas in a constant Newtonian gravity for simplicity.
In a different context, the effect of modifying the theory of gravity can be expressed as an effective change in the EOS~\cite{nojiri}.

For a static and spherically symmetric Newtonian star of radius $R$, the mass profile is obtained by integrating the balance equation
\begin{equation} \label{structure}
d P(r) = - \rho(r) g(r) dr ,
\end{equation}
where $P(r)$, $\rho(r)$, and $g(r) = GM(r)/r^2 = 4\pi G \,r^{-2} \int^r_0 dr'{r'}^2\rho(r') dr'$ are the pressure, the density, and the gravitational field at $r$, respectively.
Because Eq.~\eqref{structure} is a differential equation for two unknown functions $P(r)$ and $\rho(r)$, we need an additional equation relating the two unknowns, {\it e.g.}, the so-called EOS.
To see how we obtain this relation, let us assume that the star consists of an ideal gas.
For simplicity, let us ignore the effect of gravitation on these gaseous particles as is the case for most reports in the literature.
Then, a thermodynamic state of small enough volume in this star can be characterized uniquely in terms of the volume $V$, the number of gaseous particles $N$, and the temperature $T$.
The pressure $P$ in the volume becomes, from the ideal gas law,
\begin{equation} \label{EOS0}
P V= N k_B T,
\end{equation}
where $k_B$ is the Boltzmann constant.
Equation~\eqref{EOS0} is not appropriate for integrating Eq.~\eqref{structure} until additional constraints are given relating the three independent thermodynamic variables $V$, $N$, and $T$.
Assume that the small volume contains statistically a sufficient number of particles and does not exchange particles with the environment, {\it i.e.}, $\delta N=0$.
We further assume that the volume is adiabatic, {\it i.e., } $\delta S=0$.
 Then, Eq.~\eqref{EOS0} gives $P dV + V dP = N k_B dT$, and
the first law of thermodynamics, $dU = T dS -P dV$, gives $dT = - P dV/C_V $, with the definition of heat capacity, $dU = C_V d T$.
Combining these two equations, one gets
\begin{equation} \label{EOS:111}
\Big( 1+\frac{Nk_B}{C_V} \Big)  P dV + V dP = 0 .
\end{equation}
Finally, this gives the relation between the pressure and density as
\begin{equation} \label{EOS:1}
 P = K \rho^{\gamma},
\quad
\end{equation}
where $K$ is an integration constant, $ \gamma = (C_V+N k_B)/C_V$, and $\rho\equiv M/V$ with $M$ being the total mass of an infinitesimal volume.
For the case of a monatomic gas, $\gamma=5/3$.
Equation~\eqref{EOS:1} can be regarded as a local EOS that is satisfied by $P(r)$ and $\rho(r)$ at each spacetime point.
Equation~\eqref{structure} now becomes complete with this additional relation, Eq.~\eqref{EOS:1}, and can be integrated.

\section{THERMALLY DISCONNECTED SYSTEM IN CONSTANT GRAVITY}
\label{sec2}

Suppose that gravity's effect has not been ignored.
The presence of a gravitational field actually redistributes the particles so that the EOS obtained above might be modified accordingly.
To address this issue, let us consider a system of $N$ identical particles with mass $\mu_0$ in a box that is small enough relative to the size of the star so that the gravity does not vary much inside.
Let the height and the bottom area of the box be $2L $ and $A$, respectively.
We assume that the number of particles is statistically sufficient and that the particles
are in thermal equilibrium at temperature $T$.
Let a constant gravitational field $g$ be acting toward the negative $z$-direction.
The Hamiltonian of a particle is then given by
\begin{equation} \label{H1}
H=\frac{1}{2} \mu_0 v^2+ \mu_0 g z, \qquad -L\leq z \leq L,
\end{equation}
where the origin of the gravitational potential is set to $z=0$.
On the whole, the system can be treated as a canonical ensemble of ideal gas.
Parts of the topic dealt with here were studied in Ref.~\cite{Landsberg:1994}, and the  relativistic version is given in Ref.~\cite{LouisMartinez:2010nh}.
Theoretical consideration of matter states in the presence of gravity is addressed in Refs.~\cite{Martinez:1996vy,Martinez:1996,Sorkin:1981}.
We follow the notations in Ref.~\cite{Landsberg:1994} except for the range of $z$, which was taken to be $0\leq z\leq L$.
Let us summarize the results in Ref.~\cite{Landsberg:1994} briefly.
The canonical partition function for $N$-particles is
\begin{equation} \label{ZN}
\log Z_N = N \log Z_1 -\log N!\,,
\end{equation}
where the one-particle partition function is
\begin{equation} \label{Z1}
\log Z_1 \equiv \log \left[\left(\frac{\mu_0}h\right)^3
	\int_V d^3 x \int d^3v \, e^{-\beta H} \right]
= \log \frac{V}{(\hbar/(\mu_0 c))^3}
    	+ \frac32 \log\frac{k_B T}{2\pi \mu_0c^2}+\log \frac{\sinh X}{X}.
\end{equation}
Here, the volume of the box, $V= 2L \times A$, and the thermal energy of one degree of freedom, $k_B T$, are divided by the corresponding natural units.
The characteristic of gravity relative to kinetic energy is
\begin{equation} \label{X:def}
X  \equiv \frac{M\G}{Nk_B T}
= \frac{Mg L}{Nk_B T} = \frac{\mu_0 c^2}{k_B T} \times \frac{gL}{c^2} \geq 0; \qquad
\G \equiv g L, \quad M=N\mu_0.
\end{equation}
Now, the state of the system is characterized by four parameters: $N$, $V$, $T$, and $X$.
The energy of the system is
\begin{equation} \label{UN}
U(T,X) \equiv - \left[\frac{\partial \log Z_N}{\partial \beta}\right]_V =
	\left(\frac{5}2 - X\coth (X)  \right)N k_B T .
\end{equation}
This value decreases from $U(T,0)= \frac32 N k_B T$ to $ U(T,\infty) \to -\infty$ monotonically.
The entropy $S$ of the system, by using Stirling's approximation, is
\begin{eqnarray} \label{SN}
\frac{S}{Nk_B} \equiv \frac{U}{Nk_B T}
                  + N^{-1}\log Z_N(X)
  =\frac72+\log \frac{V/N}{(\hbar/\mu_0 c)^3}
        + \frac32 \log\frac{k_B T}{2\pi \mu_0 c^2}
           + \log \big(\frac{ \sinh X}{X}\big) - X\coth X.
\end{eqnarray}
The heat capacity
\begin{equation} \label{CV}
C_{V}\equiv \frac{\partial U}{\partial T}
= C_{V,0}+ Nk_B \left (1-\frac{X^2}{\sinh^2X} \right)
\end{equation}
varies from $C_{V,0}$ to $C_{V,0}+Nk_B$ with increasing $X$, where $C_{V,0}$ denotes the heat capacity in the absence of the gravity.
For a monatomic gas, $C_{V,0}=3Nk_B/2$.

A new point, which we would like to address, is that not only the kinetic energy, $K=3Nk_B T/2$, but also the gravitational potential energy,
\begin{equation} \label{Omega}
\Omega =U - K=
	N k_B T(1- X \coth X),
\end{equation}
contributes to the total energy.
Therefore, one may define a heat capacity for constant $T$, measuring the change of the gravitational potential energy with respect to an increment in $\G$, as
\begin{equation} \label{CG}
\frac{C_{T}}{M} \equiv \frac1{M}\frac{\partial \Omega}{\partial \G} = \frac{X}{\sinh^2X}- \coth X.
\end{equation}
The capacity for gravitational potential energy takes non-positive values to represent the attractive property of gravity and monotonically decreases from $0$ to $-1$ with increasing $X$.

The distribution of the particles depends on $z$ because the gravity is pulling down the particles to the bottom of the box.
The local distribution of the particles is described by the number density per unit phase volume $d^3 r d^3v$ at $(\vec r,\vec v)$ as
\begin{equation} \label{n(z,v)}
n(z,v) = \frac{N}{Z_1} \left(\frac{\mu_0}{h}\right)^{3} e^{-\beta H}.
\end{equation}
Integrating over the velocity, we find that
the number density per unit volume at height $z$ is given by
\begin{equation}\label{n}
n(z) \equiv \int d^3v\, n(z,v) = \frac{N}{V}\frac{X}{\sinh X }
	\, e^{-\beta \mu_0 g z} .
\end{equation}
Evaluating the momentum transfer per unit time and unit area, we find that the pressure is isotropic and is given by
\begin{eqnarray} \label{pz}
P(z) &\equiv& \int_{v_z>0} d^3 v (2p_z) v_z n(z,v)
  =  \frac{N k_B T}{V} \frac{X}{\sinh X} e^{-\frac{\mu_0 gz}{k_B T}}= n(z) k_B T .
\end{eqnarray}
From this,  we find that the average pressure
\begin{equation} \label{PbarV}
\bar P\equiv \frac{1}{2L} \int_{-L}^{L} P(z) dz = \frac{Nk_B T}V
\end{equation}
satisfies the same relation as the conventional EOS in Eq.~\eqref{EOS0} with the replacement $P \to \bar P$.
Later in this work, we use $\bar PV$ in place of $N k_B T$ from time to time.

Note that the characteristic $X$ presents
the pressure difference as the weight per unit area:
\begin{equation} \label{DP}
\Delta P \equiv P(L)-P(-L)= -2 X \bar P = - \frac{Mg}{A},
\end{equation}
which is nothing but the discrete generalization of the balance equation in Eq.~\eqref{structure}.
In the limit $L\to 0$, Eq.~\eqref{DP} reproduces Eq.~\eqref{structure}.
The gravitational potential in Eq.~\eqref{Omega} can also be obtained from $\Omega \equiv \int_V d^3 x (\mu_0 g z) n (z) $.

Because the system interacts with the gravity, the first law of thermodynamics should include the effects of gravity.
Let us derive the first law starting from the definition of the entropy, $S/k_B \equiv U/k_B T + \log Z_N$.
Differentiating both sides, we get
\begin{equation} \label{dU}
  dU = T dS + U \frac{dT}{T} - k_B T \,d\log Z_N.
\end{equation}
Differentiating Eq.~\eqref{Z1}, we can represent $d\log Z_N$ in terms of $d\beta$, $dV$, and $d\G$.
Then, the first law of thermodynamics becomes
\begin{equation} \label{dU:SVG}
dU =T dS - \bar P dV + \Omega (d\log\G).
\end{equation}
An important message in this equation is that the gravitational potential energy contributes to the first law of thermodynamics.

For an adiabatic system, we find a {\it new} EOS that describes the dependence of $\bar P$ both on $V$ and $\G$.
From the definitions of the heat capacity, Eq.~\eqref{CV}, and the gravity capacity, Eq.~\eqref{CG}, we have
$
dU = C_{V} dT + C_{T} d\G
	=\frac{ C_{V}}{Nk_B} (V d\bar P + \bar P dV) + C_{T} d\G ,
$
where we use Eq.~\eqref{PbarV} in the second equality.
By using the first law, Eq.~\eqref{dU:SVG}, for an isentropic process with $d S=0$, we have
$$
\frac{ C_{V}}{Nk_B} (V d\bar P + \bar P dV)
	+ \G\, C_{T} (d\log\G )
= - \bar P dV + \Omega (d\log\G).
$$
By using Eqs.~\eqref{CV}, \eqref{CG}, \eqref{PbarV} and $
d\log \G = d\log (\bar P V X/M )= \frac{d\bar P}{\bar P} + \frac{dV}{V}+ \frac{dX}{X},
$
we can simplify this equation to
\begin{equation} \label{eos1}
\frac{3}2 \frac{d\bar P}{\bar P} + \frac52 \frac{dV}{V} = \left(1-\frac{X^2}{\sinh^2X}\right)\frac{dX}{X}.
\end{equation}
Equation~\eqref{eos1} is integrable to give the {\it new} EOS for the gas as
\begin{equation} \label{eos2}
\bar P
   =\bar K(X) \bar \rho^{5/3};
   \qquad
\bar K\equiv
    K \left(\frac{X e^{X\coth(X)-1}} {\sinh X}\right)^{2/3}.
\end{equation}
The dependences of the pressure on $\bar \rho\equiv M/V$ and $X$ appear to be separable.
However, this does not imply that gravity's contribution is separable from the density contribution because
\be{X}
X= \frac{M\G}{Nk_B T} = \frac{MgL}{\bar P V} = \frac{\bar\rho gL}{\bar P}
	= \frac{Mg/A}{2\bar P}
\ee
contains not only the gravity but also the thermodynamic parameters,\footnote{The last term is most useful for understanding the physical implication of $X$. Namely, $X$ denotes the ratio between the weight per unit area and the average pressure.} where we use Eq.~\eqref{PbarV} in the second equality.
Therefore, the {\it new} EOS for a finite-size system is different from the conventional EOS in Eq.~\eqref{EOS:1} in the presence of a gravity.
To find the average pressure, we need to know $X$ in addition to the density $\bar\rho$.
Now, the adiabatic ideal gas is uniquely characterized when two of the three quantities $\bar \rho$, $\bar P$ and $X$ are identified.
Especially, the pressure difference from the top to the bottom comes from Eq.~\eqref{DP} once $X$ is given.

Let us illustrate the {\it new} EOS in several limiting cases.
For the case $M\G \ll N k_B T$ (i.e., $X\ll 1$), we have
$$
\bar P \approx K \bar \rho^{5/3} \Big(1+ \frac19 X^2 + \cdots \Big);
\qquad X \equiv \frac{\G}{K \bar \rho^{2/3}}.
$$
Notice that the correction in the $O(X)$ does not appear probably because thermodynamic property would be invariant under a change of gravity's direction.
When one deals with an astrophysical object, one may regard a macroscopic object as an assembly of many subsystems, each having a small enough height while containing a sufficient number of particles to be dealt with statistically.
As long as this is allowed, $X ~(\propto L)$ for a given subsystem can be chosen to be small enough by taking its height to be very small.
Therefore, for most cases, taking the $L\to 0$ limit of the subsystem is possible.
Then, gravity's effect on the EOS for the subsystem is ignorable.
In this sense, the conventional EOS in Eq.~\eqref{EOS:1} for an adiabatic system is locally reliable even when gravity's effect has been considered.

When $X\gg 1$, on the other hand, the {\it new} EOS becomes
\be{EOS:large}
\bar P \approx {K}^{3/5} \left(\frac{2 M\G}{e}\right)^{2/5} \bar \rho^{7/5}.
\ee
This case includes a strong gravity region ($X\propto g$).
The Newtonian framework holds only when the particles move much slower than light.
If a particle freely falls from the center of the box to the bottom, its velocity approaches the velocity of light when $gL \approx c^2$.
Therefore, $X \ll \mu_0 c^2/k_B T$ is required to remain in the non-relativistic region.
For low temperature, $X$ is still allowed to be large within the Newtonian regime if
$k_B T < \mu_0 gL \ll \mu_0 c^2$.
In this case, the size effects in the EOS in Eq.~\eqref{EOS:large} will be observable.
Explicit examples of this type are matters during the radiation-dominated period of the universe and in the cores of extremely dense stars~\cite{Chavanis:2002,Lynden-Bell}.

Once we know that the {\it new} EOS in Eq.~\eqref{eos2} shows a characteristic difference from that of the conventional one, we can display how to organize the structure of a spherically symmetric star of radius $R$.
We assume that the star is composed of many small systems, each containing a statistically sufficient number of particles $N_i$ and occupying the smallest possible volume $V_i$.
We divide the radius into many pieces and label each of them with an integer $i$.
The outermost slice is labeled by the number $0$, and the number increases for inner slices.
Now, the radial size of a system $\delta r_i = V_i^{1/3} \ll R$ will depend on the number density.
As seen in Eq.~\eqref{eos2}, gravity's effect on a system's EOS is ignorable if
$$
\delta r_i  \ll \delta r_c \equiv \frac{2k_B T_i}{\mu_0 g_i},
$$
 where $T_i$ and $g_i$ are the temperature and the gravity at $r_i$, respectively.
For a high-temperature system with a weak gravity, $\delta r_c$ is very large.
Physical quantities such as the gravity, the density, and the pressure are almost constant over the volume $V_i$.
In this case, Eqs.~\eqref{DP} and \eqref{eos2} simply reproduce the balance equation, Eq.~\eqref{structure}, and the conventional EOS, Eq.~\eqref{EOS:1}, respectively, in the limit $\delta r_i \to 0$.
The balance equation can be integrated conventionally.

In the presence of a strong gravity at a low temperature, on the other hand, $\delta r_c$ can be comparable or smaller than $\delta r_i$.
In this case, as seen in Eq.~\eqref{DP}, the pressure changes drastically in $V_i$.
When an ordinary polytropic star is considered, this happens at the surface of the star where the pressure gradient is non-vanishing, $\nabla P \neq 0$, but the average pressure becomes zero, $\bar P \to 0$.
Therefore, around the surface of the star, the EOS in Eq.~\eqref{eos2} does not go to the EOS in Eq.~\eqref{EOS:1}.
In addition, the pressure change, Eq.~\eqref{DP},
\be{DP2}
\Delta P_i = -2X_i \bar P_i
\ee
over $\delta r_i$ is not infinitesimal, but takes the form of a discrete difference equation over the radial slices, which is a generalization of Eq.~\eqref{structure}.
In fact, the thicknesses of the slices experiencing this drastic change in pressure is extremely thin relative to the radius.
To show this, we sum Eq.~\eqref{DP2} over $i$ from the star's surface inward.
Consider a slice of thickness $\delta r_0 \ll R$, an average density $\rho(R)$, an average pressure $p(R)$, and a pressure difference $\Delta P_0/P_0 = 2X_0 \gg 1$.
The number of particles in a volume $V_0$ is given by $N_0 = \rho(R)/\mu_0 \times V_0$, which is comparable to Avogadro's number.
We assume that each small system contains the same number of particles, i.e., $N_i = N_0$.
For an ordinary hard star such as a neutron star or a white dwarf, this condition restricts the thickness $\delta r_0$ to be of the order of a centimeter, which is much smaller than their radii.
The number density of particles in the volume $V_1$ in the slice just inside of $V_0$ is increased by $\Delta n/\bar n = -2X_0$.
This gives
 $
 n_{1} \equiv n_{R-\delta r_0} = (2X_0 +1) n_0.
 $
Therefore, the number density at the slice located at $r=R-\delta r_0$ is increased by the factor $2X_0+1$.
Because the number of particles in $V_1$ is the same as that in $V_0$, the volumes are related by
$V_1 \equiv (\delta r_1)^3 = V_0/(2X_0 +1)$.
With this choice of volume, the value of $X_1\equiv X_{R-\delta r_0}$ is decreased to
$$
X_{1} \approx \frac{X_0}{(2X_0 +1)^{1/3}},
$$
where we have assumed that the values of $g$ and $T$ are almost the same as those in $V_0$.
If this relation is continued inside, $X_{20} \approx 0.1$ and $X_{150} \approx 0.01$ for almost all values of $X_0> 1$.
This implies that $\delta r_{i} \ll \delta r_c$ if $i> 150$.
Therefore, for the slices inside this radius, one may use Eq.~\eqref{EOS:1} safely.
The sum of the thicknesses corresponding to the slices from $i=0$ to $i=150$ is
$$
\Delta r_{0 \sim 150} \equiv \sum_{i=0}^{150} \delta r_i = \sum_{i=1}^{150} \frac{X_i }{X_0} \times \delta r_0 \approx 1.34 \delta r_0
$$
for $X_0 = 100$.
Therefore, we conclude that the thickness having nontrivial $X$ is of $O(\delta r_0)$, which is negligible compared to the radius of the star.
Inside the thin layer from the star's surface, the conventional EOS can be used to integrate the balance equation.
Therefore, the effect of the modified EOS on the structure of a star will be negligibly small.

\section{SUMMARY AND DISCUSSION}
\label{sec3}
We have examined whether or not a (local) conventional EOS obtained in theories without gravity is credible in a physical situation with gravity.
We have found that the conventional EOS is reliable as long as the weight of the box per unit area is much smaller than the average pressure.
When one deals with an astrophysical object, one may regard a macroscopic object as an assembly of many subsystems, each having a small enough size while containing a statistically sufficient number of particles.
Therefore, we can use the conventional EOS safely to integrate the balance equation.
Even though this is true for most astrophysical systems, exceptions for which the small size limit of Eqs.~\eqref{DP} and \eqref{eos2} may not be applicable for various reasons are possible.
For example, the size of a system can be restricted to be larger than the size of its elements.
If the element has a macroscopic size, the box should also be chosen to be macroscopic however dense the elements are.
Examples are a scalar dark matter model~\cite{Sin:1992bg,Lee:1995af} and the case of a cosmic Hawking radiation~\cite{Parikh:2002qh,Lee:2008vn}.
In these cases, the de Broglie wavelengths of the elements are long.
Because of the long wavelength, the quantum properties and self-gravity of the particles will be important.

This topic requires additional studies based on quantum theory.
Another example is the case where the spacetime curvature (or gravity) is extremely large around a region (or point) so that a statistically sufficient number of particles should occupy the volume satisfying $X  \gtrsim 1$.
Examples are as follows:
\begin{enumerate}
\item The (self-)gravity increases equally or faster than $L^{-1}$ as the system size $L$ decreases. This happens near the surface of a star or during phase transitions in the theories of Palatini gravity~\cite{Barausse:2007ys,Kim:2013nna}. Even though the present results are based on Newtonian gravity, they indicate the importance of gravitational corrections. If the precise form of the EOS is to be determined, detailed additional analyses are necessary based on  the corresponding modified gravity theories.

\item The gravity just above the event horizon of a black hole diverges\footnote{The divergence of the acceleration can be understood from the resemblance between the black hole horizon and the Rindler horizon. The acceleration of a Rindler observer becomes infinite as he approaches the horizon.} to a locally static observer. In this case, the small size limit may not be applicable however small the system is. A non-negligible effect will be present on the EOS of matter around there. A general relativistic treatment is required.

\end{enumerate}
For case 1, the origin of the pressure difference could be the self gravity, which is non-negligible compared to the background gravity.
Therefore, knowledge of how to deal with the self-gravity effect is important.

With respect to an observer who cognizes only the directions orthogonal to the gravity,
the {\it new} EOS, Eq.~\eqref{eos2}, presents an interesting gravity effect.
One may rewrite Eq.~\eqref{eos2} as
\be{eos:br}
\bar P(x) = \sigma ^{5/3}(x) \bar K(\frac{ g \sigma(x)}{ 2\bar P(x)} );
 \quad \bar K(y) = \frac{ K(y)}{(2L)^{5/3}},
\ee
where $\sigma \equiv  M/A $ is the surface mass density, $x$ represents the coordinates for the orthogonal directions, and $\bar P$ plays the role of the pressure along the orthogonal directions.
Explicitly, if an observer lives on a brane world and particles are restricted to remain in a narrow region around the brane due to gravity, the EOS on the brane takes the form of Eq.~\eqref{eos:br}.

\section*{ACKNOWLEDGMENT}
This work was supported by a grant from the 2014 program for visiting professors overseas at the Korea National University of Transportation.
GK was supported in part by the Academic Program of Asia-Pacific Center for Theoretical Physics (APCTP) and the R\&D Program of Korea Institute of Science and Technology Information (KISTI).
HK thanks Chueng Ji for his hospitality.


\begin{thebibliography}{99}
\bibitem{Lattimer:2012nd}
  J.~M.~Lattimer,
  Ann.\ Rev.\ Nucl.\ Part.\ Sci.\  {\bf 62}, 485 (2012)
  [arXiv:1305.3510 [nucl-th]].

\bibitem{cosmology}
V. Mukhanov, {\it Physical Foundation of Cosmology} (Cambridge University Press, Cambridge, England,  2012).

\bibitem{Bonnor}
W. B. Bonnor,
Mon. Not. R. Astr. Soc. {\bf 116}, 351 (1956).

\bibitem{Lombardi:2001ms}
  M.~Lombardi and G.~Bertin,
  Astron.\ Astrophys.\  {\bf 375}, 1091 (2001)  
  [astro-ph/0106336].

\bibitem{adam}
C. Adam, C. Naya, J. Sanchez-Guillen, R. Vazquez, and A. Wereszczynski,
Phys. Rev. C {\bf 92}, 025802 (2015) [arXiv:1503.03095];
Phys. Lett. B {\bf 742}, 136 (2015) [arXiv:1407.3799].

\bibitem{Kim:2013nna}
  H-C.~Kim,
  Phys.\ Rev.\ D {\bf 89}, 064001 (2014)
  [arXiv:1312.0705 [gr-qc]].


\bibitem{Banados:2010ix}
  M.~Banados and P.~G.~Ferreira,
  Phys.\ Rev.\ Lett.\  {\bf 105}, 011101 (2010)
  [arXiv:1006.1769 [astro-ph.CO]].


\bibitem{Barausse:2007ys}
  E.~Barausse, T.~P.~Sotiriou and J.~C.~Miller,
  Class.\ Quant.\ Grav.\  {\bf 25}, 105008 (2008)
  [arXiv:0712.1141 [gr-qc]].

\bibitem{Sotiriou:2008rp}
  T.~P.~Sotiriou and V.~Faraoni,
  Rev.\ Mod.\ Phys.\  {\bf 82}, 451 (2010)
  [arXiv:0805.1726 [gr-qc]].

\bibitem{Olmo:2011uz}
  G.~J.~Olmo,
  Int.\ J.\ Mod.\ Phys.\ D {\bf 20}, 413 (2011)
  [arXiv:1101.3864 [gr-qc]].

\bibitem{Pani:2012qd}
  P.~Pani and T.~P.~Sotiriou,
  Phys.\ Rev.\ Lett.\  {\bf 109}, 251102 (2012) 
  [arXiv:1209.2972 [gr-qc]].

\bibitem{nojiri}
S. Nojiri 
 and S. D. Odintsov,
Phys. Rept. {\bf 505}, 59 (2011)
[arXiv:1011.0544 [gr-qc]];
S. Capozziello, 
S. Nojiri, 
and S. D. Odintsov, 
Phys. Lett. B {\bf 634}, 93 (2006)
[hep-th/0512118].

\bibitem{Landsberg:1994}
P. T. Landsberg, J. Dunning-Davies, and D. Pollard,
Am.\ J.\ Phys.\ {\bf 62}, 712 (1994).
\bibitem{LouisMartinez:2010nh}
  D.~J.~Louis-Martinez,
  Class.\ Quant.\ Grav.\  {\bf 28}, 035004 (2011)
  [arXiv:1012.3063 [physics.class-ph]].

\bibitem{Sorkin:1981}
R. D. Sorkin, R. M. Wald, and Z. Z. Jiu,
 Gen. Rel. Grav. {\bf 13}, 1127 (1981).

\bibitem{Martinez:1996}
E. A. Martinez,
Phys.\ Rev.\ D {\bf 53} 7062 (1996).

\bibitem{Martinez:1996vy}
  E.~A.~Martinez,
  Phys.\ Rev.\ D {\bf 54}, 6302 (1996)
  [gr-qc/9609048].


\bibitem{Chavanis:2002}
P. H.~Chavanis, A\&A {\bf 381}, 709 (2002).

\bibitem{Lynden-Bell}
D. Lynden-Bell and R. Wood,
Mon. Not. R. Astr. Soc. {\bf 138}, 495 (1968).

\bibitem{Lee:1995af}
  J.~W.~Lee and I.~G.~Koh,
  Phys.\ Rev.\ D {\bf 53}, 2236 (1996)
  [hep-ph/9507385].
\bibitem{Sin:1992bg}
  S.~J.~Sin,
  Phys.\ Rev.\ D {\bf 50}, 3650 (1994)
  [hep-ph/9205208].

\bibitem{Parikh:2002qh}
  M.~K.~Parikh,
  Phys.\ Lett.\ B {\bf 546}, 189 (2002)
  [hep-th/0204107].

\bibitem{Lee:2008vn}
  J.~W.~Lee, H.~C.~Kim and J.~Lee,
  Mod.\ Phys.\ Lett.\ A {\bf 25}, 257 (2010)
  [arXiv:0803.1987 [hep-th]].
\end{thebibliography}
\end{document}